\documentclass{moriond}

\usepackage{amsmath}
\usepackage{amssymb}

\newcommand{\Fermi}{{\textit{Fermi}}}

\newcommand{\psize}{0.49}

\def\be{\begin{equation}}
\def\ee{\end{equation}}
\def\bea{\begin{eqnarray}}
\def\eea{\end{eqnarray}}

\newcommand{\ben}{\begin{enumerate}}
\newcommand{\een}{\end{enumerate}}
\newcommand{\bi}{\begin{itemize}}
\newcommand{\ei}{\end{itemize}}

\newcommand{\dmgen}{
\cite{2011PhLB..697..412H,2011PhRvD..84l3005H,2012PhRvD..86h3511A,2013PhRvD..88h3521G,2013PDU.....2..118H,2015JCAP...03..038C,2016PDU....12....1D,2017ApJ...840...43A}}

\newcommand{\mspgen}{
\cite{2011JCAP...03..010A,2012PhRvD..86h3511A,2013PhRvD..88h3521G,2015JCAP...02..023P,2015ApJ...812...15B}}

\newcommand{\stat}{
\cite{2015JCAP...05..056L,2016PhRvL.116e1102B,2016PhRvL.116e1103L,2019PhRvL.123x1101L,2020PhRvL.125l1105L,2020PhRvL.124w1103Z,2020PhRvL.125x1102L,2021PhRvL.127p1102C,2022PhRvD.105f3017M,2023JCAP...06..013C,2024arXiv240204733M}}

\newcommand{\pop}{\cite{2015ApJ...812...15B,2016JCAP...08..018H,2018MNRAS.481.3966B,2020JCAP...12..035P,2022JCAP...06..025D}}

\newcommand{\Photo}{\includegraphics[height=35mm]{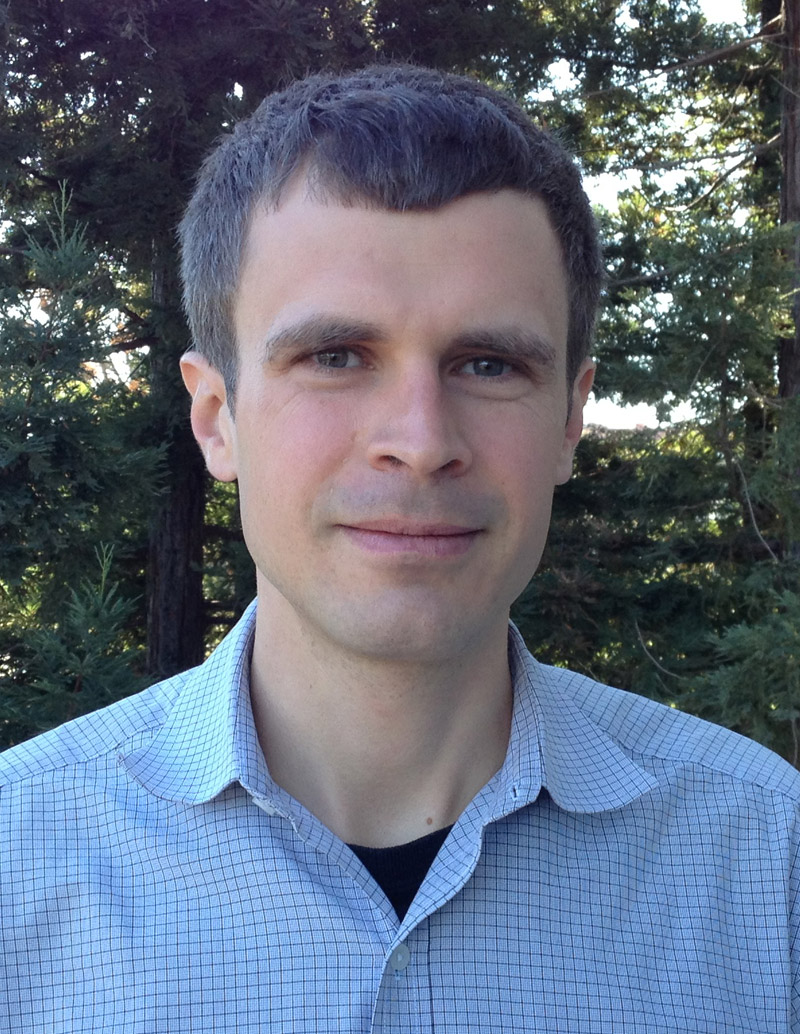}}

\begin{document}

\vspace*{4cm}
\title{Galactic center GeV excess and classification of \Fermi-LAT sources with machine learning}

\author{ Dmitry V. Malyshev }

\address{Erlangen Centre for Astroparticle Physics, \\
Nikolaus-Fiebiger-Str. 2, Erlangen 91058, Germany}


\maketitle\abstracts{
Excess of gamma rays with a spherical morphology around the Galactic center (GC) observed in the {\Fermi} large area telescope (LAT) data is one of the most intriguing features in the gamma-ray sky. 
The excess has been interpreted by annihilating dark matter as well as emission from a population of unresolved millisecond pulsars (MSPs). 
We use a multi-class classification of \Fermi-LAT sources with machine learning to study the distribution of MSP-like sources among unassociated \Fermi-LAT sources near the GC. 
We find that the source count distribution of MSP-like sources is 
comparable with the MSP explanation 
of the GC excess.
}

\section{Introduction}

\begin{figure}
\begin{minipage}{\psize\linewidth}
\centerline{\includegraphics[width=0.99\linewidth]{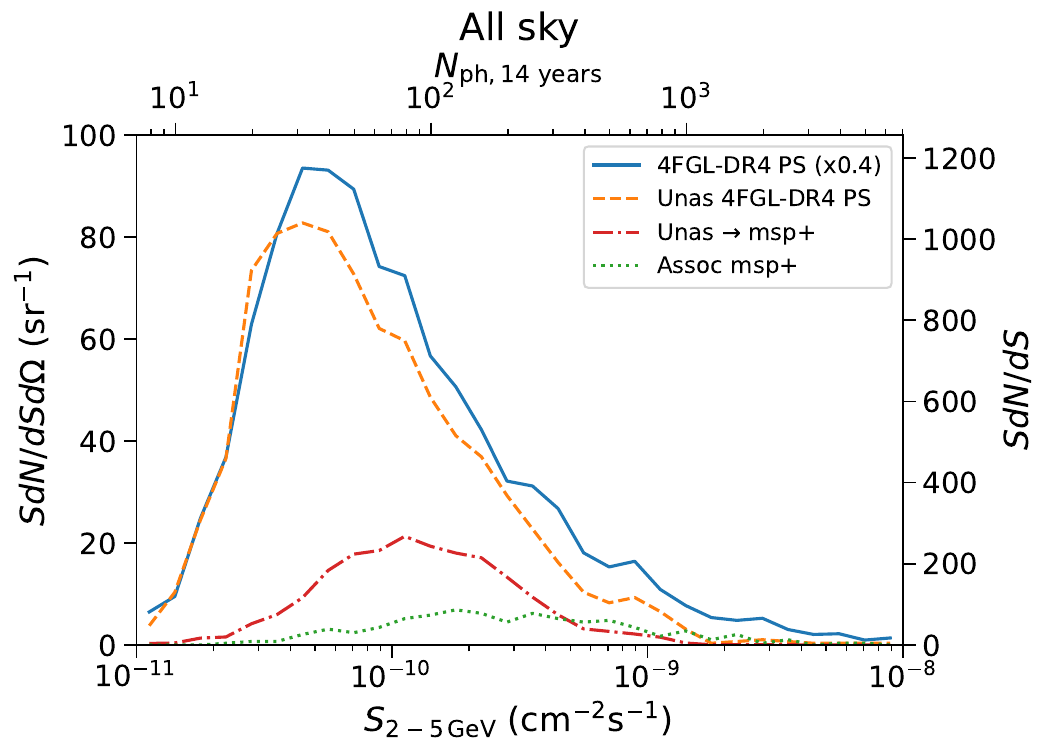}}
\end{minipage}
\hfill
\begin{minipage}{\psize\linewidth}
\centerline{\includegraphics[width=0.99\linewidth]{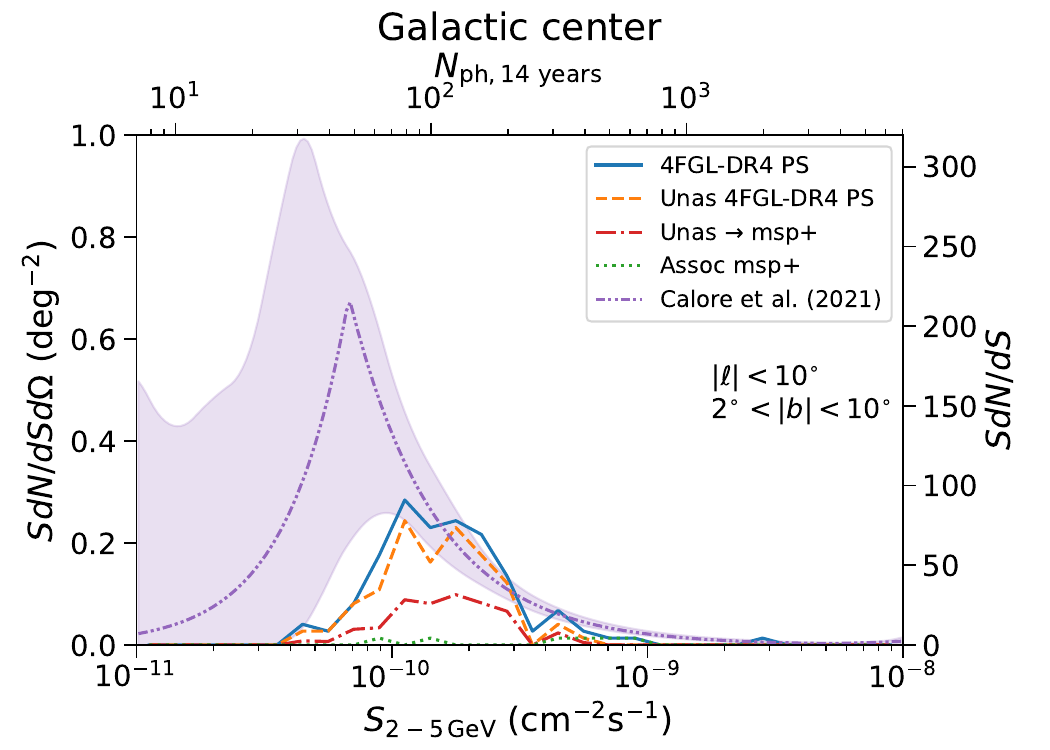}}
\end{minipage}
\caption[]{
Source count distributions of 4FGL-DR4~\cite{2023arXiv230712546B} sources for the whole sky (left panel) and in the GC ROI (right panel).
Blue solid lines --  all sources, orange dashed lines -- unassociated sources, green dotted lines -- associated MSP-like sources, 
red dash-dotted lines -- estimated source count for MSP-like sources among unassociated sources.
Purple dash-dot-dotted line (right panel) -- estimated source count distribution from photon count statistics~\cite{2021PhRvL.127p1102C}.
}
\label{fig:dNdS}
\end{figure}

Excess of gamma rays with GeV energies near the GC observed in the \Fermi-LAT data has sparked a significant interest as it has spectral and spatial distributions expected for dark matter (DM) annihilation\dmgen.
Nevertheless, there also exist astrophysical explanations of the excess, e.g., due to a population of unresolved MSPs\mspgen.
Approaches to test the MSP hypothesis can be separated into two general classes: (1) population studies based on physical modeling or on the extrapolation from associated MSPs and globular clusters~\pop; (2) statistical methods that exploit the properties of distributions of gamma rays near the GC~\stat.
Advantages and challenges in these two classes of methods are:
\ben
\vspace{-2mm}
\item {\bf Population studies}\\
{\bf Pros:} The methods are based on observed properties of local MSPs and MSPs in globular clusters and can be used to separate the contribution of MSPs near the GC from contributions of other sources.\\
{\bf Cons:} Extrapolation from the distribution of nearby MSPs or observed globular clusters is needed.
\item {\bf Statistical methods}\\
{\bf Pros:} Sensitive to sources both above and below the detection threshold. It is possible to determine sources correlated with a particular distribution, e.g., spherical profile around the GC.\\
{\bf Cons:} Not specific to MSPs, i.e., only the overall distribution of sources is determined in a relatively large energy bin.
\een
We illustrate the challenges in the population studies in Fig.~\ref{fig:dNdS}. 
On the left panel we show the distribution of 
4FGL data release 4 (DR4) catalog~\cite{2023arXiv230712546B}
source counts as a function of flux in energy bin between 2 and 5 GeV.
The corresponding number of photons in this energy range for 14 years of data taking is shown on the top x-axis.
It is estimated using the acceptance of 2.5 $\rm m^2 sr$ around a few GeV reduced by 20\% to account for
dead time (mostly due to passage of South Atlantic Anomaly)\footnote{\url{https://www.slac.stanford.edu/exp/glast/groups/canda/lat_Performance.htm}}. 
On average, the fluxes of associated MSPs and globular clusters (green dotted line) are larger than the fluxes of unassociated sources (orange dashed line).
In order to determine the expected distribution of MSPs near the GC one needs to model spatial distribution of MSPs, time evolution, and environment.
For instance, the density of stars is important for estimates of the rate of creation of binaries, where MSPs can be produced.

As an example of statistical methods, we compare the estimates of the source count distribution based on photon counts statistics~\cite{2021PhRvL.127p1102C}
with the 4FGL-DR4 catalog source counts  in Fig.~\ref{fig:dNdS}, right panel.
The analysis of ref.~\cite{2021PhRvL.127p1102C} uses 12 years of data in energy range 2 -- 5 GeV with $|\ell| < 10^\circ$, $2^\circ < |b| < 10^\circ$ (referred as GC ROI below).
The 4FGL-DR4 catalog is based on 14 years of data. We calculate the flux of sources in the 2 -- 5 GeV energy bin using log-parabola parametrization of the spectrum.
There is a very good agreement between the photon counts statistics method and the actual counts of sources with fluxes $S_{\rm 2 - 5\, GeV} \gtrsim 10^{-10}\ {\rm cm^{-2}\, s^{-1}}$.
For smaller fluxes, the detection efficiency in the \Fermi-LAT catalog decreases, while the photon counts method predicts more sources than are detected in the catalog.
Although the photon counts statistics method gives a reasonable estimate of the total distribution of sources, it cannot separate MSPs from other types of sources, it also has large uncertainty for $S_{\rm 2 - 5\, GeV} < 10^{-10}\ {\rm cm^{-2}\, s^{-1}}$.

One of the caveats of the MSP hypothesis has been a lack of associated MSPs near the GC~\cite{2016JCAP...08..018H,2017JCAP...05..056H}.
Indeed, there are only 6 associated MSPs and globular clusters in the GC ROI (shown by the green dotted line in the right panel of Fig.~\ref{fig:dNdS}).
We note, however, that most of sources in the vicinity of the GC are unassociated (dashed orange line in the right panel of Fig.~\ref{fig:dNdS})
and there can be a significant number of MSPs among them.

\section{Classification of sources using machine learning}

Machine learning (ML) methods have been used to probabilistically separate different classes of sources and, in particular, to determine the expected contribution from MSP-like sources 
among unassociated sources
near the GC~\cite{2024arXiv240104565M}.
ML methods enable one to determine the expected contribution from MSP-like sources from the gamma-ray data near the GC, thus reducing the necessity to perform extrapolation from the distribution of nearby MSPs in population studies.
Moreover, ML methods provide a separation of MSP-like sources from the other source types near the GC, which is very difficult to do with statistical methods based on photon count maps.

The contributions of MSP-like sources among unassociated sources over the whole sky (in the GC ROI) obtained with random forest classification~\cite{2024arXiv240104565M}
is shown by the red dash-dotted line on the left (right) panel in Fig.~\ref{fig:dNdS}.
The classification is performed with seven input features describing spectral properties of sources, detection significance, and significance of variability, and five output classes dominated BL Lacs, FSRQs, pulsars, pulsar wind nebulae + supernova remnants, and by MSPs~\cite{2024arXiv240104565M}.
The class dominated by MSPs (denoted as msp+ below) also contains globular clusters and normal galaxies~\cite{2024arXiv240104565M}. 
Since there are only 6 galaxies detected, their contribution to this class is insignificant.
Overall, in the GC ROI there are 118 sources, among which there are 91 unassociated sources and 6 MSPs or globular clusters. The expected number of msp+ sources is 38.8, 
i.e., more than 6 times larger than the number of associated msp+ sources and more than two times smaller than the number of unassociated sources in this ROI.

\begin{figure}
\begin{minipage}{\psize\linewidth}
\centerline{\includegraphics[width=0.99\linewidth]{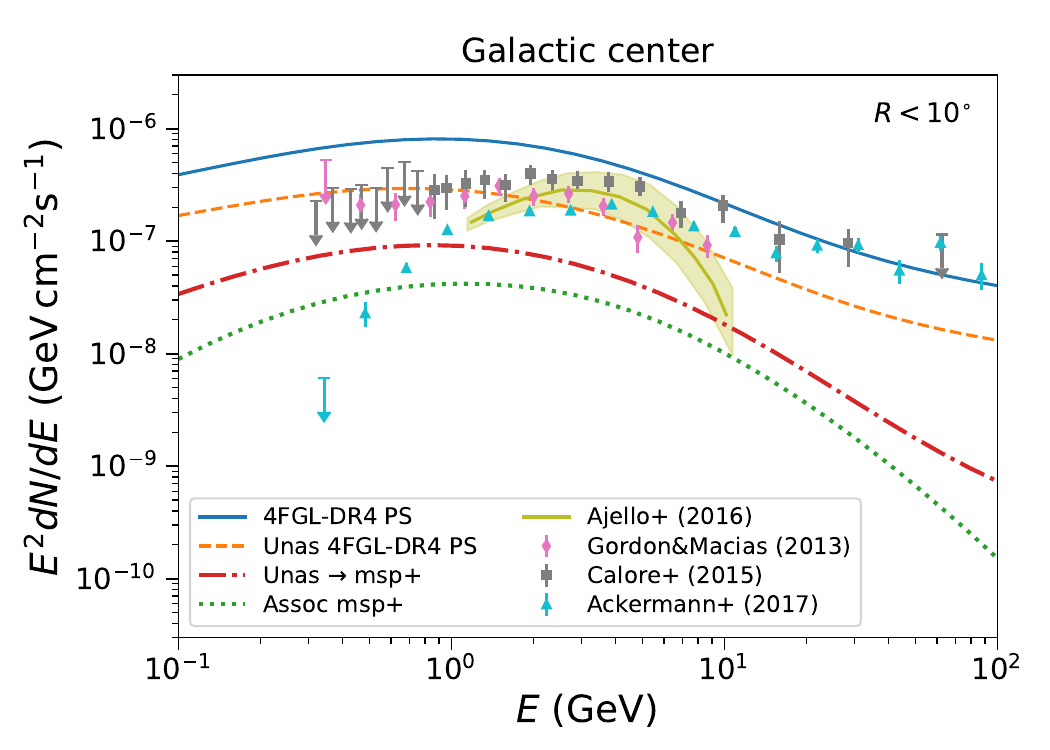}}
\end{minipage}
\hfill
\begin{minipage}{\psize\linewidth}
\centerline{\includegraphics[width=0.99\linewidth]{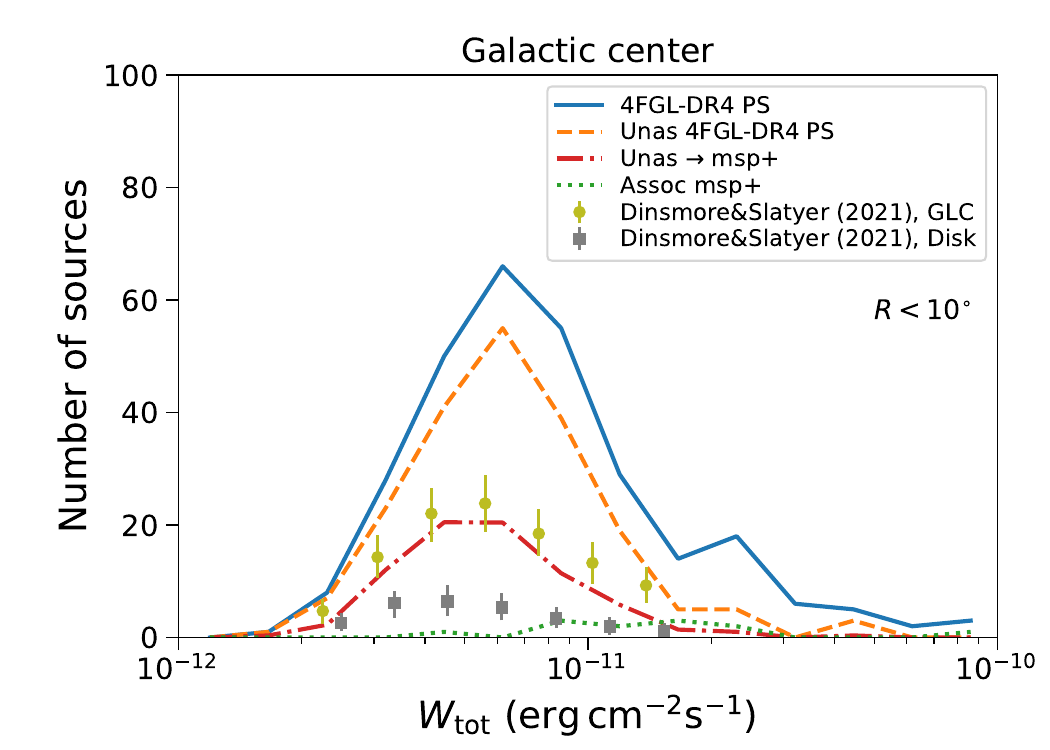}}
\end{minipage}
\caption[]{Spectral energy distributions (left panel) and source count distributions as a function of energy flux above 100 MeV (right panel) for 4FGL-DR4 sources within $10^\circ$ from the GC. 
The GCE spectra on the left panel are take from refs.~\cite{2013PhRvD..88h3521G,2015JCAP...03..038C,2017ApJ...840...43A,2016ApJ...819...44A}.
The source count distributions in the right panel are adapted from ref.~\cite{2022JCAP...06..025D}.
The lines are defined in Fig.~\ref{fig:dNdS}.}
\label{fig:spectrum}
\end{figure}

The combined spectral energy distribution (SED) of msp+ sources within $10^\circ$ from the GC 
is shown by red dash-dotted line on the left panel of Fig.~\ref{fig:spectrum}~\cite{2024arXiv240104565M}.
Here we also plot the SED of unassociated (all) 4FGL-DR4 sources, shown by the dashed orange (solid blue) line.
The SED of associated MSPs and globular clusters in this ROI is shown by the green dotted line.
The msp+ SED is about a factor of three smaller than the Galactic center excess (GCE) SED around a few GeV.
However, one should also take into account MSPs below the detection threshold.
On the right panel of Fig.~\ref{fig:spectrum} we compare the source count distribution of msp+ sources as a function of the total energy flux above 100 MeV 
(red dash-dotted line)~\cite{2024arXiv240104565M} 
with models of the source count distribution of MSPs, which are compatible with the GCE.
In particular, the source count distribution of MSP-like sources determined with ML is compatible with the population model of MSPs near the GC based on observed MSPs in globular clusters \cite{2016JCAP...08..018H,2022JCAP...06..025D} (shown by yellow circles).
We note that the expected number of MSP-like sources in this ROI (75.6) is significantly larger than the number of associated MSP-like sources (12) shown by the green dotted line
and smaller than the overall number of sources (285), blue solid line, or the number of unassociated sources (198), orange dashed line.
The estimate of the source count distribution of msp+ sources provides an additional constraint on the population models of MSP-like sources near the GC, e.g., it favors a model based on globular clusters (orange circles)~\cite{2016JCAP...08..018H} over a model based on the distribution of MSPs in the Galactic disk (gray squares)~\cite{2018MNRAS.481.3966B}.

\section{Conclusions}

In this note we show that multi-class classification of \Fermi-LAT sources with ML methods can be used to bridge the gap between models of MSP-like sources near the 
GC based on population studies and the models of distributions of sources based on photon counts statistics.
We show that the estimated number of MSP-like sources among unassociated \Fermi-LAT sources in the ROI is compatible with the 
MSP explanation of the GCE in models that predicts relatively many detected MSPs near the GC~\cite{2016JCAP...08..018H,2016PhRvL.116e1103L} rather than models dominated by MSPs below the detection threshold~\cite{2018MNRAS.481.3966B,2020JCAP...12..035P,2022NatAs...6..703G}.

\section*{Acknowledgments}

The work is supported by the DFG grant MA 8279/3-1.

\section*{References}

\bibliography{msp_multi_gce_bibl}

\end{document}